\documentclass[11pt]{article}
\pdfoutput=1
\usepackage[utf8]{inputenc}
\usepackage{amsmath}
\usepackage{amssymb}
\usepackage{siunitx}
\usepackage{graphicx}
\usepackage{slashed}
\usepackage{array}
\usepackage[leftcaption,raggedright]{sidecap}
\usepackage{caption}
\usepackage{wrapfig}
\usepackage{xcolor} 
\usepackage{booktabs}
\definecolor{myblue}{RGB}{49, 88, 154}

\usepackage[hidelinks]{hyperref}
	\hypersetup{
		colorlinks   = true, 
        linkcolor    = myblue, 
        citecolor= myblue, 
		urlcolor= black,
		}

\usepackage[natbib=true,sorting=none,backrefstyle=none, style = phys, biblabel=brackets]{biblatex}
\DeclareSIUnit{\GeV}{\giga\electronvolt}
\DeclareSIUnit{\MeV}{\mega\electronvolt}
\DeclareSIUnit{\keV}{\kilo\electronvolt}
\DeclareSIUnit{\us}{\micro\second}
\DeclareSIUnit{\ns}{\nano\second}
\DeclareSIUnit{\Hz}{\hertz}

\usepackage[top=2cm, bottom=2cm, left=2.2cm, right=2.2cm]{geometry}

\newcommand\pubnumber{NuPhys2023-Vanessa Cerrone}
\newcommand\pubdate{\today}

\def\napoli{Dipartimento di Fisica e Astronomia dell'Universit\`{a} di Padova \\ INFN Sezione di Padova, Padova, Italy}

\def\support{\footnote{vanessa.cerrone@pd.infn.it}}

\def\Title#1{\begin{center} {\Large #1 } \end{center}}
\def\Author#1{\begin{center}{ \sc #1} \end{center}}
\def\Address#1{\begin{center}{ \it #1} \end{center}}

\newcommand\pubblock{\rightline{\begin{tabular}{l} \pubnumber\\
         \pubdate  \end{tabular}}}
\newenvironment{Abstract}{\begin{quotation}  }{\end{quotation}}
\newenvironment{Presented}{\begin{quotation} \begin{center} 
             PRESENTED AT\end{center}\bigskip 
      \begin{center}\begin{large}}{\end{large}\end{center} \end{quotation}}

\def\beq{\begin{equation}}
\def\eeq#1{\label{#1}\end{equation}}
\def\eeqn{\end{equation}}

\def\beqa{\begin{eqnarray}}
\def\eeqa#1{\label{#1}\end{eqnarray}}
\def\eeqan{\end{eqnarray}}

\let\bar=\overbar

\def\Dslash{\not{\hbox{\kern-4pt $D$}}}
\def\dslash{\not{\hbox{\kern-2pt $\del$}}}

\def\msb{{\bar{\ssstyle M \kern -1pt S}}}

\usepackage{xspace}

\newcommand*\Edep{\ensuremath{E_\text{dep}}\xspace}%
\newcommand*\Evis{\ensuremath{E_\text{vis}}\xspace}%

\begin{document}
\begin{titlepage}
\pubblock

\vfill
\Title{Probing Neutrino Oscillations with Reactor Antineutrinos in JUNO}
\vfill
\Author{ Vanessa Cerrone\support, on behalf of the JUNO collaboration}
\Address{\napoli}
\vfill
\begin{Abstract}
The Jiangmen Underground Neutrino Observatory (JUNO) is a multi-purpose neutrino experiment currently under construction in South China, in an underground laboratory with approximately 650~m of rock overburden (1800 m.w.e.). The detector consists of a 20 kton liquid scintillator target, contained inside a 35.4-meter-diameter spherical acrylic vessel. The central detector (CD) is equipped with 17,612 20-inch and 25,600 3-inch Photomultipliers Tubes (PMTs), providing more than 75\% total photocathode coverage. \\
JUNO's main goal is the determination of the neutrino mass ordering with reactor antineutrinos, emitted from two adjacent nuclear power plants on a $\sim$~52.5 km baseline from the experimental site.
JUNO's strategic location at a baseline corresponding to the first solar oscillation maximum, where the kinematic phase $\Delta_{21} \simeq \frac{\pi}{2}$, grants it the unique capability to  simultaneously probe the effects of oscillations on both solar and atmospheric scales; moreover, it stands out as the first experiment to address the unresolved NMO question through vacuum-dominant oscillations.
The oscillated energy spectrum in JUNO changes subtly depending on the neutrino mass ordering, which manifests as an energy-dependent phase shift, thus providing sensitivity to this parameter. 
Furthermore, the unparalleled size and energy resolution will enable to achieve a sub-percent precision on three parameters:  $\Delta m_{21}^{2}$, $\Delta m_{31}^{2}$, and $\sin^2\theta_{12}$.\\
This contribution will focus on JUNO's oscillation physics potential with reactor antineutrinos, with a particular emphasis on its crucial role in inaugurating a new era of precision within the neutrino sector.

\end{Abstract}
\vfill
\begin{Presented}
NuPhys2023, Prospects in Neutrino Physics\\
King's College, London, UK,\\ December 18--20, 2023
\end{Presented}
\vfill
\end{titlepage}
\def\thefootnote{\fnsymbol{footnote}}
\setcounter{footnote}{0}

\section{Introduction}
The standard three-neutrino paradigm provides a well-established framework for describing neutrino oscillations~\cite{bib:pmns, bib:pdg}, and it is characterized by six (eight) parameters: three mixing angles ($\theta_{12}$, $\theta_{23}$, and $\theta_{13}$), one Dirac CP phase ($\delta_{\rm CP}$), (two additional Majorana phases), and two distinct mass squared differences ($\Delta m_{21}^2$ and $\Delta m_{31}^2$, or equivalently $\Delta m_{32}^2$). Despite this understanding, crucial aspects of neutrino properties, such as their particle nature (Dirac or Majorana), the potential violation of CP symmetry
\begin{wrapfigure}[14]{r}{0.48\textwidth}
\vspace{-0.4cm}
\centering
\includegraphics[width=0.48\textwidth]{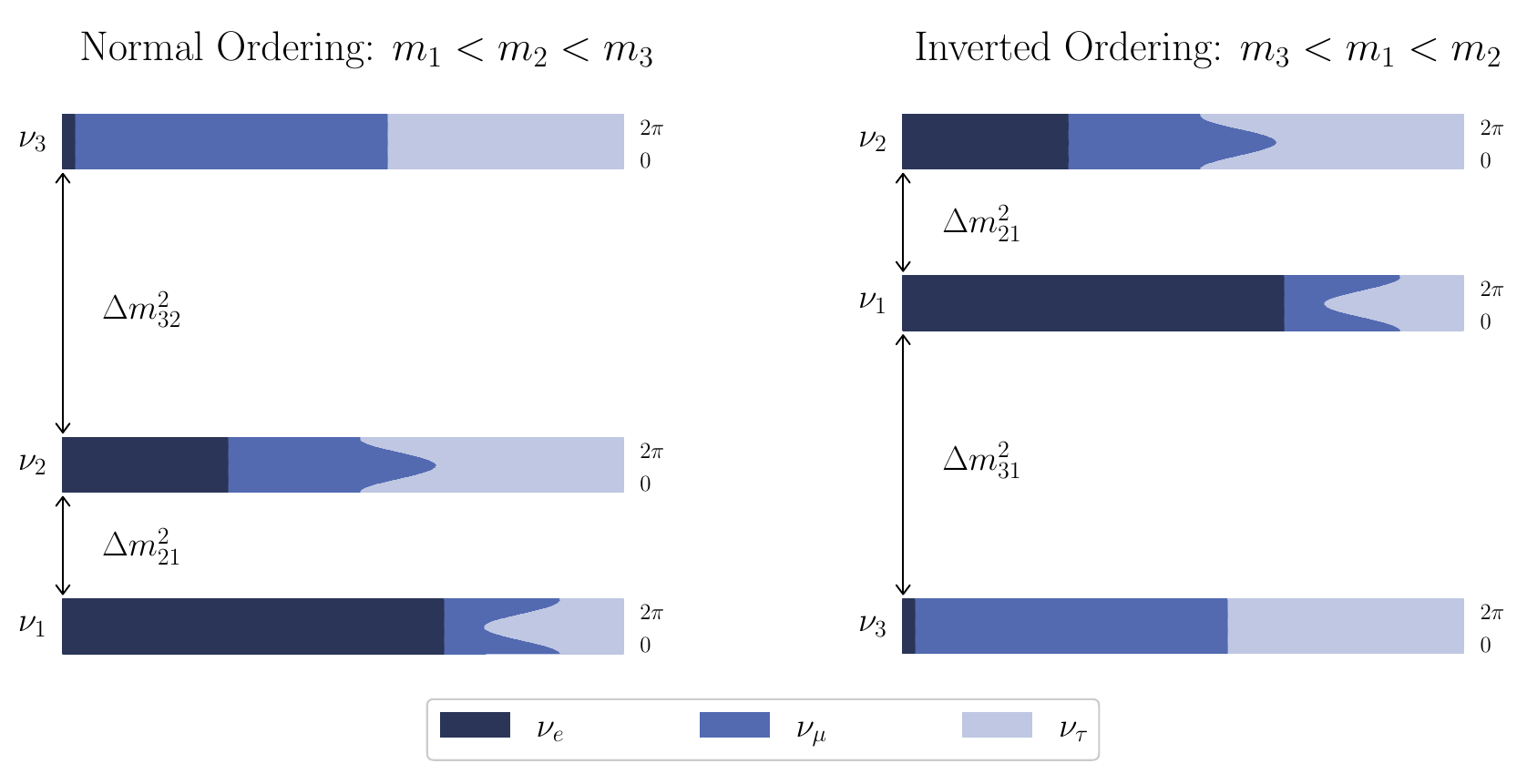}
\captionof{figure}{Visual representation of neutrino flavor ($\nu_{\alpha}$) content of each neutrino mass eigenstate ($\nu_{i}$) given the global best-fit values of the oscillation parameters, varying the CP violating phase between $0$ and $2\pi$. The two mass ordering schemes are illustrated. Inspired by \cite{bib:graphics}. }
\label{fig:scheme}
\end{wrapfigure}
  in the leptonic sector, and the arrangement of neutrino mass eigenstates (Neutrino Mass Ordering, NMO, illustrated in \autoref{fig:scheme}), remain elusive~\cite{bib:giunti}. Over the past few decades, various experiments have leveraged both natural and man-made sources to enhance our comprehension of neutrino properties and interactions. Among these, reactor antineutrinos have played a pivotal role in the landscape of neutrino physics since their initial detection~\cite{bib:reines-cowan}. 
Reactor antineutrinos constitute a pure and intense source of electron antineutrinos ($\overline{\nu}^{}_e$), with energies reaching approximately \SI{10}{\MeV}, contrasting with the GeV scale of muon neutrinos and antineutrinos generated in beams from accelerators. 
The Jiangmen Underground Neutrino Observatory (JUNO)~\cite{bib:junophysics} is a multi-purpose 20~kton liquid scintillator (LS) experiment currently under construction in South China. JUNO is primarily designed for the determination of the NMO exploiting electron antineutrinos, emitted from six 2.9~GW$_{\rm th}$ and two 4.6~GW$_{\rm th}$ reactor cores in the Yangjiang and Taishan nuclear power plants (NPPs), respectively. A satellite detector, called Taishan Antineutrino Observatory~\cite{bib:tao} (TAO or JUNO-TAO), will be deployed at a distance of roughly \SI{40}{\meter} from one of the Taishan reactors. The major goal of TAO is to provide a model-independent
and data-driven reference reactor spectrum for JUNO. The location of both JUNO and TAO is shown in the map on the left side of \autoref{fig:map-osc}. The driving requirements for the design of the JUNO detector, dictated by its ambitious physics goals, include an energy resolution within 3\% at \SI{1}{\MeV}, a precise control of the energy scale
(overall non-linearity effects below 1\%), and a substantial antineutrino statistics~\cite{bib:junophysics}. A standout feature is the unparalleled total photo-coverage of more than 75\%, granted by a dual photo-detection system, comprising 17,612 20-inch PMTs and 25,600 3-inch PMTs. More information on the detector can be found in~\cite{bib:junophysics}.

\section{Oscillation physics with reactor antineutrinos in JUNO}
\begin{figure}[t]
    \centering
    \includegraphics[width=0.48\textwidth]{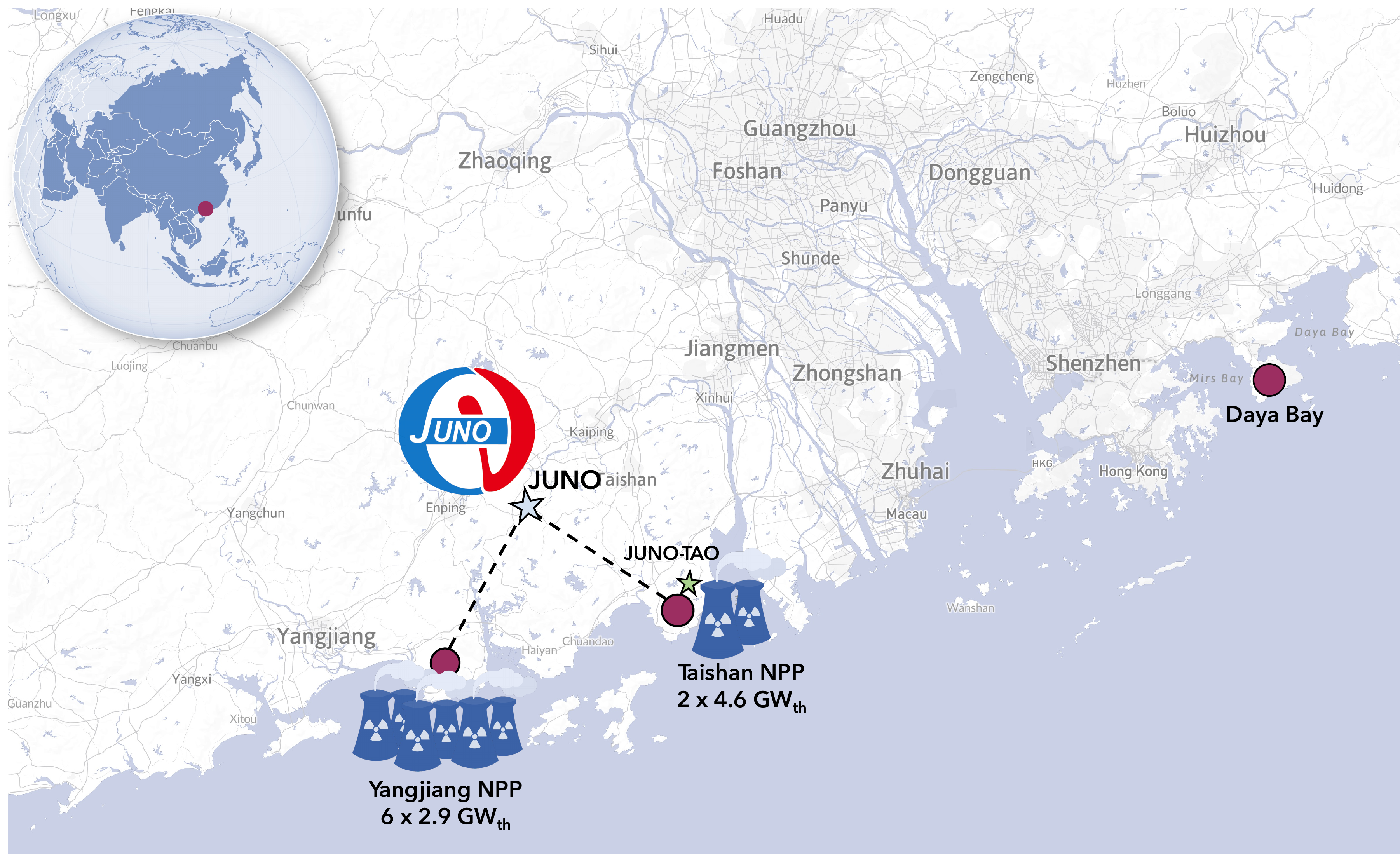}
   \hspace{0.5cm}
 \includegraphics[width=0.45\textwidth]{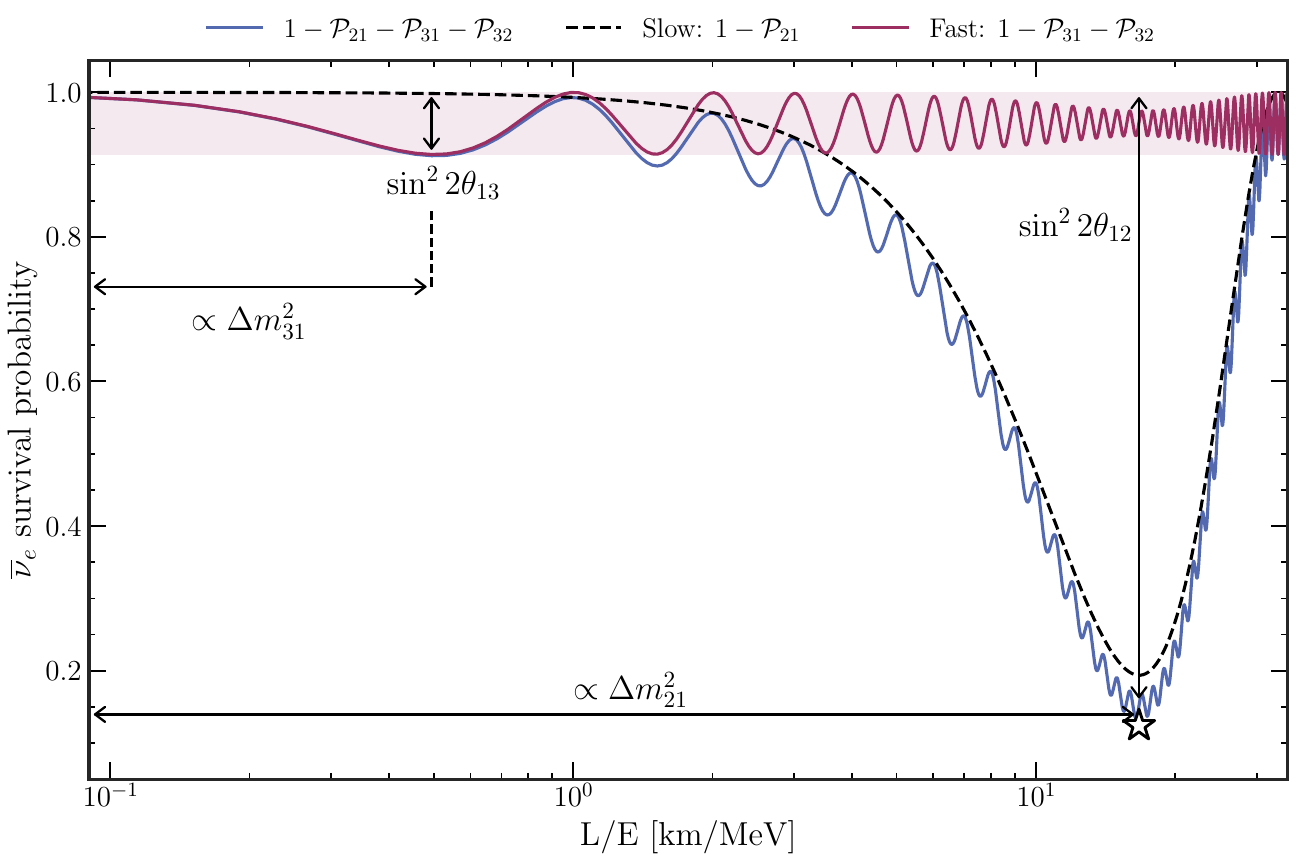}
    \caption{Left: Location of JUNO and TAO. Right: Electron antineutrino survival probability according to \autoref{eq:nue_survival_vacuum} as a function of $L/E$. The solar (slow) and atmospheric (fast) oscillations are represented by the black dashed line and the solid red line, respectively. }
    \label{fig:map-osc}
\end{figure}

The experiment is located at a baseline $L$ of approximately 
52.5~km, sitting in the first solar oscillation maximum, i.e., where the kinematic phase $\Delta_{21} \equiv \frac{\Delta m^2_{21} L}{4E} \simeq \frac{\pi}{2}$. This medium baseline configuration allows to exploit three generation effects and measure four oscillation parameters with a single experiment. Indeed, JUNO stands out as the first experiment to simultaneously probe vacuum-dominant oscillations on both the solar and atmospheric scales~\cite{bib:msw}. Benefiting from a pure $\overline{\nu}^{}_e$ source, JUNO is sensitive to the electron antineutrino survival probability, which has the following expression (in vacuum)~\cite{bib:giunti}:
\begin{eqnarray}
{\cal P}(\overline{\nu}^{}_e \to \overline{\nu}^{}_e)
&  =  & 
1 - \hspace{0.05cm} \sin^2 2\theta_{12}\, {c}_{13}^4\, \sin^2 \Delta_{21}
-  \sin^2 2\theta_{13}
\left(  c_{12}^2\sin^2 \Delta_{31} +  s_{12}^2 \sin^2 \Delta_{32}\right) \\
\nonumber
&  =  &  1 - \hspace{0.05cm} {\cal P}_{21}- {\cal P}_{31}- {\cal P}_{32}   ,
\label{eq:nue_survival_vacuum}
\end{eqnarray}

where $c_{ij} \equiv \cos \theta_{ij}$, $s_{ij} \equiv \sin \theta_{ij}$, $\Delta_{ij}={\Delta m^2_{ij} L}/{4E}$, and ${\cal P}_{ij}$ represent the three terms associated with the respective $\Delta_{ij}$-induced oscillations. Notably, there is no dependence on $\theta_{23}$ or $\delta_{\rm CP}$. \autoref{fig:map-osc} (right) illustrates ${\cal P}(\overline{\nu}^{}_e \to \overline{\nu}^{}_e)$ as a function of $L/E$: fast and slow components are distinguished with different colors, and JUNO's location around the first solar oscillation maximum is pointed out by the star marker. Matter effects have a minor impact compared to long-baseline oscillation experiments, yet including them in real-data analysis is crucial given the foreseen precision in the measurement of oscillation parameters~\cite{bib:matterJUNO}.

\subsection{Antineutrino interaction and detector response}
\begin{figure}[h]
    \centering
    \includegraphics[width=0.47\textwidth]{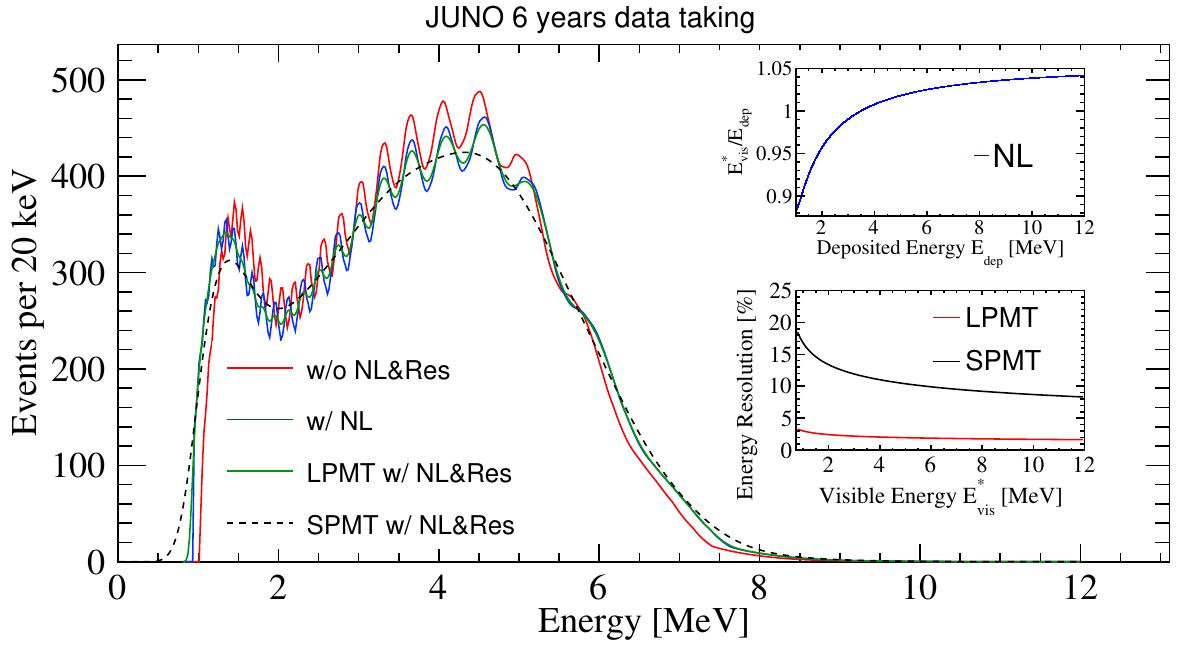}
    \hfill
 \includegraphics[width=0.48\textwidth]{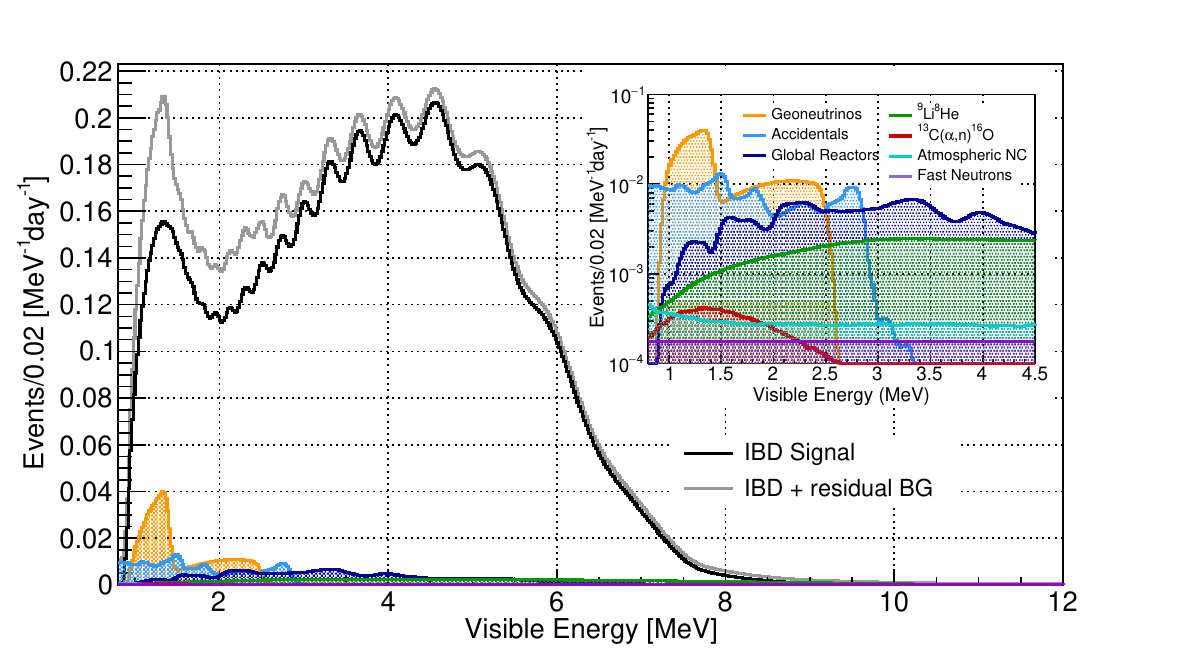}
    \caption{Left: Expected prompt energy spectrum with and without the different detector response effects, i.e., liquid scintillator non-linearity (NL) and energy resolution (Res). Right: Expected energy spectra in JUNO after selection, with all residual spectral components. Taken from \cite{bib:sub_osc}.}
    \label{fig:detectorresponse_bkg}
\end{figure}

JUNO detects reactor antineutrinos through the Inverse Beta Decay (IBD) reaction: 
\begin{equation*}
    \overline{\nu}^{}_e+p \rightarrow e^{+}+n~,
\end{equation*}
which is considered the golden channel for their detection thanks to its well-understood and precisely predicted cross section, much larger than for other processes at this energy scale.
Upon interaction, the emitted positron rapidly releases its energy and annihilates into two \SI{0.511}{\MeV} photons, thus giving rise to a \textit{prompt} signal. Meanwhile, the neutron undergoes multiple scatterings, eventually thermalizing within the detector medium. Approximately \SI{220}{\us} later, it is predominantly captured by a proton in the Hydrogen-rich LS, emitting a \SI{2.22}{\MeV} $\gamma$-ray, thereby generating a \textit{delayed} signal. The positron retains nearly all of the incoming antineutrino kinetic energy, hence it is a reliable proxy for the latter. Consequently, the energy spectrum generated by prompt signals serves as the experimental observable for studying the $\overline{\nu}^{}_e$ oscillation pattern. \\
Accurately predicting the measured prompt spectrum requires a thorough understanding of the detector's response, namely of all the processes mapping the antineutrino energy to experimentally accessible quantities, i.e., the amount of light collected by the PMTs.
Initially, positrons interacting in the LS generate photons through scintillation and, to a lesser extent, Cherenkov radiation mechanisms. However, due to the quenching effect, not all deposited energy converts into light, resulting in a non-linear relationship between deposited energy and the number of scintillation photons detected by the PMTs. This 
feature is embedded in the Liquid Scintillator Non-Linearity (LSNL) curve, characterized by the following relation: $\Evis = f_{\mathrm{LSNL}}(\Edep) \cdot \Edep$, where \Edep is the deposited energy, \Evis is the visible energy assuming perfect energy resolution, and $f_{\mathrm{LSNL}}(\Edep)$ denotes the LSNL function. In the analysis, the Daya Bay non-linearity curves~\cite{bib:dayabayNLcurves} are adopted\footnote{This choice stems from the similarity in the LS composition between Daya Bay and JUNO.} with appropriate adjustments to ensure consistency with the latest JUNO simulations~\cite{bib:calibration}.
Furthermore, the visible energy undergoes additional smearing due to the finite energy resolution of the detector~\cite{bib:calibration}. \autoref{fig:detectorresponse_bkg} (left) illustrates the expected prompt energy spectrum in JUNO, both with and without the aforementioned detector response effects— liquid scintillator non-linearity (NL) and energy resolution (Res).

\subsection{Event selection and backgrounds}
The IBD reaction provides a distinctive double signature to achieve effective signal/background discrimination. 
Among the major backgrounds, it is possible to identify both \textit{correlated} and \textit{uncorrelated} components. 
\textit{Correlated} backgrounds arise from a single physics process: within this class, for example, geoneutrinos interact 
through IBD, thus being completely indistinguishable from signal events. Moreover, some long-lived cosmogenic isotopes, 
(mainly $^9$Li and $^8$He), undergo $\beta^-$ decay with an accompanying neutron in the final state, mimicking the IBD 
signature. Contrariwise, uncorrelated backgrounds, also called \textit{accidental coincidences}, are related to two 
independent signals passing all selection criteria by chance. \\ 
The IBD selection strategy is based upon the typical prompt-delayed pattern and is effective in drastically reducing backgrounds while retaining high efficiency for signal events. For example, a fiducial volume (FV) cut at 17.2 meters is employed to mitigate the radioactive background rate, since it exponentially increases at the edges of the target. Then, time and vertex cuts are applied according to the typical IBD $\Delta R$-$\Delta t$ signature, where  $\Delta R$ and $\Delta t$ are the time interval and the distance between prompt and delayed candidates, respectively. The energy windows are defined by $E_{\rm prompt} \in (0.7, 12.0)$~MeV for prompt events and $E_{\rm delayed} \in (1.9, 2.5) \cup (4.4, 5.5)$~MeV for delayed events.
In-depth information about backgrounds and selection criteria can be found in \cite{bib:sub_osc}. The reconstructed energy spectrum, comprising the reactor antineutrino signal and all 
residual backgrounds is reported in the right panel of \autoref{fig:detectorresponse_bkg}.

\section{Sensitivity to NMO and oscillation parameters}

\begin{figure}[t]
    \centering
    \includegraphics[width=0.5\textwidth]{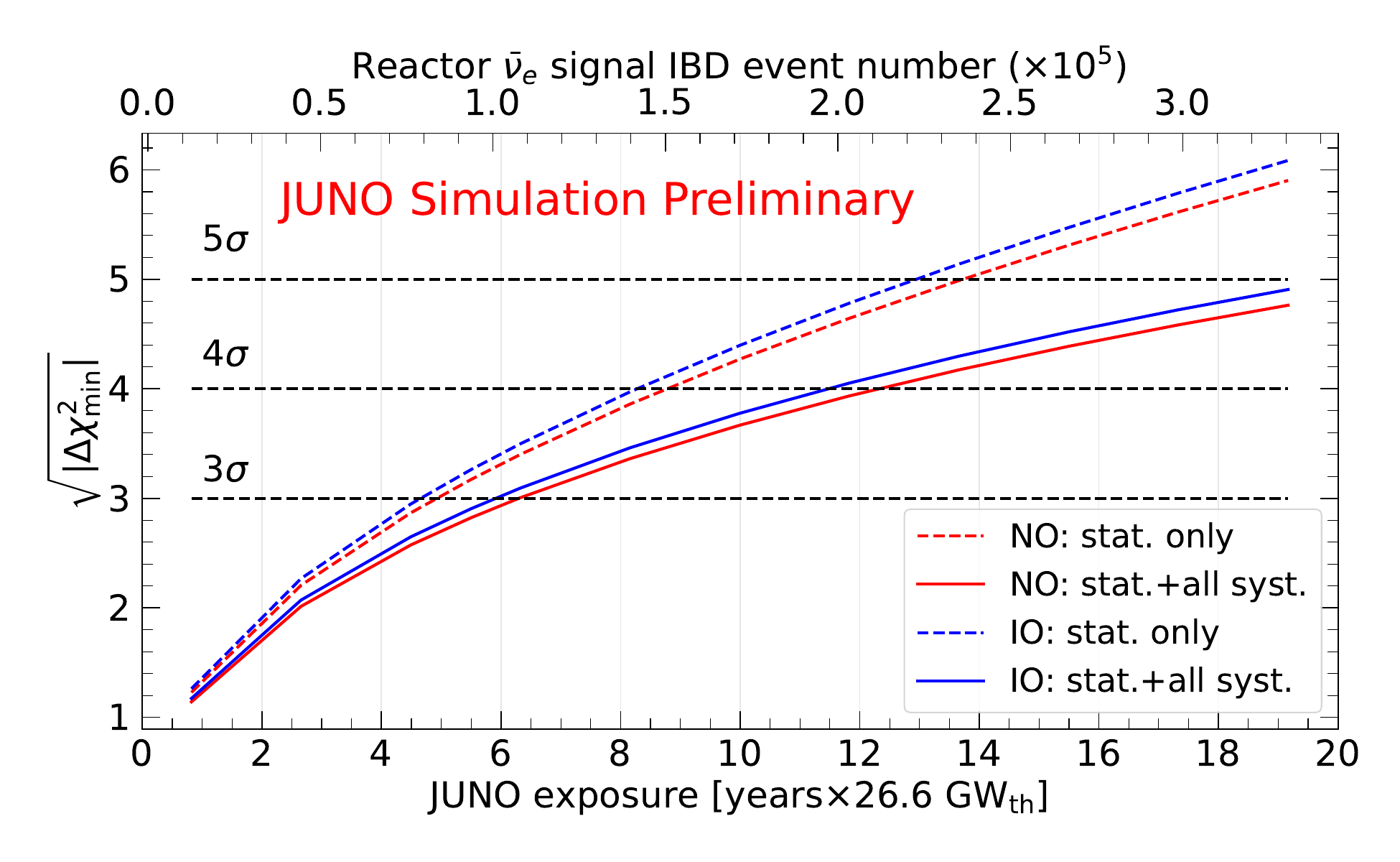}
    \hfill
 \includegraphics[width=0.45\textwidth]{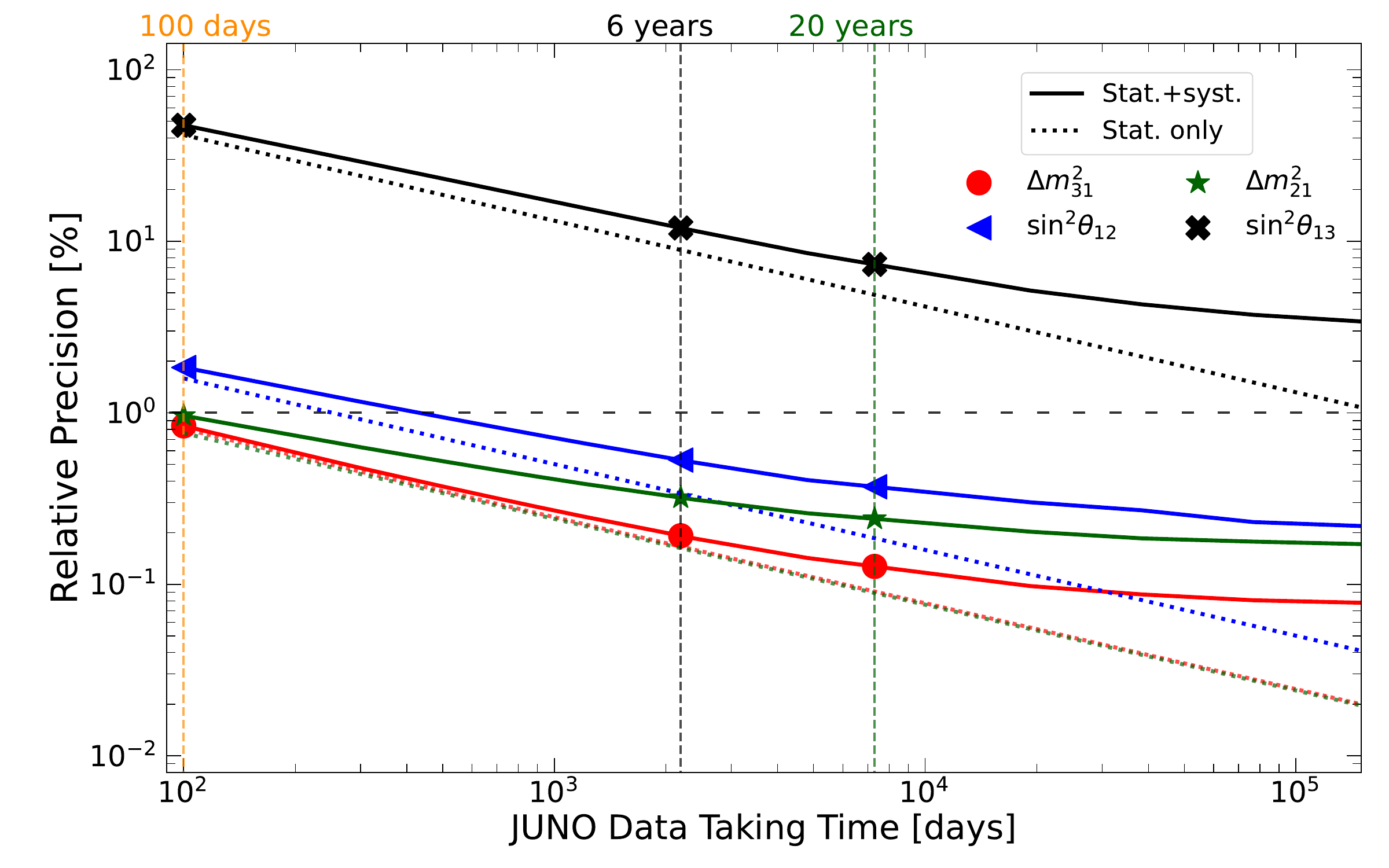}
    \caption{Left: MO median sensitivity as a function of JUNO exposure~\cite{bib:mo}. Right: Relative precision on oscillation parameters as a function of JUNO exposure~\cite{bib:sub_osc}.}
    \label{fig:nmo}
\end{figure}

To extract the neutrino oscillation parameters and assess the NMO sensitivity, the analysis involves simultaneously fitting JUNO and TAO nominal spectra, against a hypothesis model based on the standard three-flavor framework. TAO simulated data is used to constrain the reactor antineutrino energy spectrum, while the oscillation pattern is inferred from JUNO spectrum. An Asimov pseudo-dataset is built under both the Normal Ordering (NO) and Inverted Ordering (IO) hypotheses. Then, the median sensitivity discriminator is defined as:
 $   \Delta \chi^2 \equiv |\chi^2_{\rm min} (\rm NO) - \chi^2_{\rm min} (\rm IO)|. $ 
The resulting $ \Delta \chi^2 $ is reported in \autoref{fig:nmo} (left panel) as a function of JUNO data taking time for both NO (red) and IO (blue) hypotheses: with $\sim 6.2$ years of data taking at full 26.6 GW$_{\rm th}$ reactor power, JUNO can reach a median sensitivity of $\sim3\sigma$~\cite{bib:mo}. Furthermore, ongoing studies are exploring the possibility to incorporate additional information from the detection of atmospheric neutrinos~\cite{bib:junophysics, bib:mo}. \\ The obtained relative precision on the oscillation parameters~\cite{bib:sub_osc} is reported in \autoref{fig:nmo} (left) and \autoref{tab:params_table}. It is estimated that with 6 years of data, JUNO can determine the parameters $\Delta m_{21}^{2}$, $\Delta m_{31}^{2}$, and $\sin^2\theta_{12}$ with a precision of $\sim$~0.2\%, $\sim$~0.3\%, and $\sim$~0.5\%, respectively. Moreover, JUNO is foreseen to already exceed global precision on these parameters within the first months of data acquisition.

\begin{table}[h]
\centering
\begin{tabular}{cccccc}
\toprule
          & Central value  & PDG 2020~\cite{bib:pdg}   & 100 days  & 6 years & 20 years \\ \midrule
$\Delta m_{21}^2$ [eV$^2$] & \num{7.53e-05}   & 2.4\%  & 1.0\%    & 0.3\%   & 0.2\%    \\
$\Delta m_{31}^2$ [eV$^2$] & \num{2.5283e-03}    & 1.3\% &  0.8\%    & 0.2\%    & 0.1\%    \\
$ \sin^2\theta_{12}$ & 0.307      & 4.2\% &  1.9\%    & 0.5\%   & 0.3\%    \\
$ \sin^2\theta_{13}$ & 0.0218     & 3.2\%  &  47.9\% &  12.1\%  & 7.3\%    \\ \bottomrule
\end{tabular}
\captionof{table}{Current relative precision on oscillation parameters from global analysis (PDG 2020)~\cite{bib:pdg}, and JUNO (different exposures) in this analysis, assuming true NO~\cite{bib:sub_osc}.}
\label{tab:params_table}
\end{table}

\section{Conclusions}

JUNO is a next-generation liquid scintillator neutrino observatory currently under construction in South China. Thanks to its unprecedented size and expected performances, it will be able to measure $ \Delta m^2_{31} $, $ \Delta m^2_{21} $, and $ \sin^2 \theta_{12} $ with sub-percent precision, marking a milestone in the field. Moreover, JUNO stands out as the only experiment currently capable of simultaneously probing two different oscillation frequencies and resolving the NMO through vacuum-dominant oscillations of reactor antineutrinos. On top of this, the precision measurement of neutrino oscillation parameters is a  powerful tool to test the standard 3-flavor neutrino model. For example, it will allow a direct test the of the PMNS matrix unitarity (specifically of the first row) by leveraging the combination with short-baseline reactor antineutrino experiments and solar neutrino experiments, and to narrow down the parameter space of the effective mass of the neutrino-less double beta decay~\cite{bib:junophysics}. 
\newpage
\printbibliography



\end{document}